\documentclass[sigconf, screen, nonacm]{acmart}

\usepackage{multirow}
\usepackage{xspace}
\usepackage{wasysym}
\usepackage{algorithm}
\usepackage[noend]{algpseudocode}
\usepackage{enumerate}
\usepackage{enumitem}
\usepackage[bottom]{footmisc}
\usepackage{makecell}

\setcopyright{none}

\newcommand{\red}[1]{{\color{red} #1}}

\newcommand{\hide}[1]{}
\allowdisplaybreaks  % Allows page break inside align/math environments

\begin{document}

\title{Query Generation with Direct Preference Optimization for Document Expansion in E-commerce Search}
\settopmatter{authorsperrow=4}

\author{Kaihao Li}
\authornote{Contributed equally to this research.}
\affiliation{%
  \institution{Walmart Global Technology}
  \city{Hoboken}
  \country{USA}
}
\email{kaihao.li@walmart.com}

\author{Feng Liu}
\authornotemark[1]
\authornote{Work done while at Walmart.}
\affiliation{%
  \institution{Walmart Global Tech}
  \city{Sunnyvale}
  \country{USA}
}
\email{liu.f@pku.edu.cn}

\author{Juexin Lin}
\authornotemark[1]
\authornotemark[2]
\affiliation{%
  \institution{Walmart Global Technology}
  \city{Sunnyvale}
  \country{USA}
}
\email{linjuexin@gmail.com}

\author{Xunfan Cai}
\authornotemark[2]
\affiliation{%
  \institution{Walmart Global Technology}
  \city{Sunnyvale}
  \country{USA}
}
\email{ffcaizy@gmail.com}

\author{Zhen Yang}
\authornotemark[2]
\affiliation{%
  \institution{Walmart Global Technology}
  \city{Sunnyvale}
  \country{USA}
}
\email{njuyangz@gmail.com}

\author{Tony Lee}
\authornotemark[2]
\affiliation{%
  \institution{Walmart Global Technology}
  \city{Sunnyvale}
  \country{USA}
}
\email{tonyleexyz@gmail.com}

\author{Ciya Liao}
\authornotemark[2]
\affiliation{%
  \institution{Walmart Global Technology}
  \city{Sunnyvale}
  \country{USA}
}
\email{liao_ciya@yahoo.com}

% Shortened author list for header
\renewcommand{\shortauthors}{Liu et al.}

\begin{abstract}
Doc2Query, a popular document expansion technique, leverages sequence-to-sequence models to generate relevant queries, effectively addressing the "vocabulary mismatch" problem in information retrieval. However, these models often suffer from generating either hallucinations unrelated to the document or repetitive content already present in the document. Training sequence-to-sequence models to produce high-quality, novel, and relevant tokens remains a significant challenge. To address these issues, we introduce a novel approach, QGDPO, that employs direct preference optimization (DPO) to guide the generation process. We first fine-tune a base sequence-to-sequence model and subsequently utilize a relevance model to score its predictions. Based on these scores, we construct pairs of winning and losing predictions as relevance preferences for the DPO training. Furthermore, we enhance our pipeline by using the relevance model to filter out poor predictions, retaining only the most relevant generated content for indexing. QGDPO effectively eliminates 50\% of irrelevant predictions comparing against Doc2Query baselines, while the relevance filter removes an additional 14.61\%. This feature has been successfully deployed to production on Walmart.com for full traffic, with a substantial improvement in relevance and user engagement.

\end{abstract}

% CCS Concepts (Optional on arXiv preprints, retained for completeness)



\maketitle

\section{Introduction}

\label{sec:intro}

\begin{figure*}[t]

    \centering

    \includegraphics[width=0.98\textwidth,height=0.22\textheight]{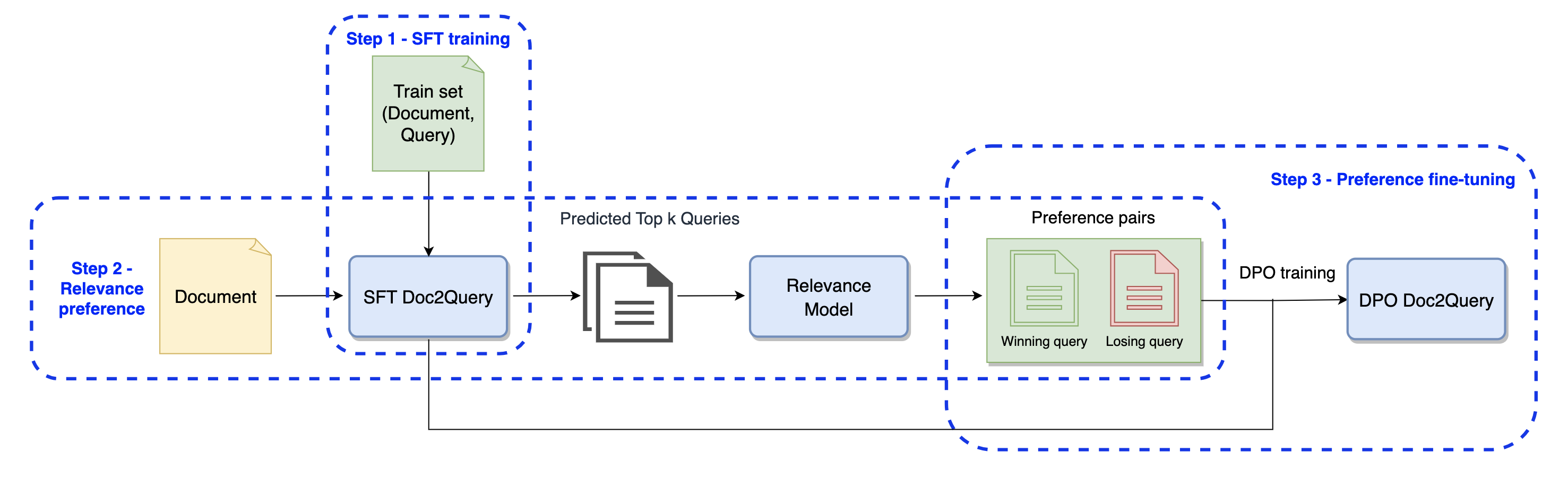}

    \caption{Training flow}

    \label{figure:dpo_train}

\end{figure*}

\begin{figure}[t]

    \centering

    \includegraphics[width=0.48\textwidth,height=0.14\textheight]{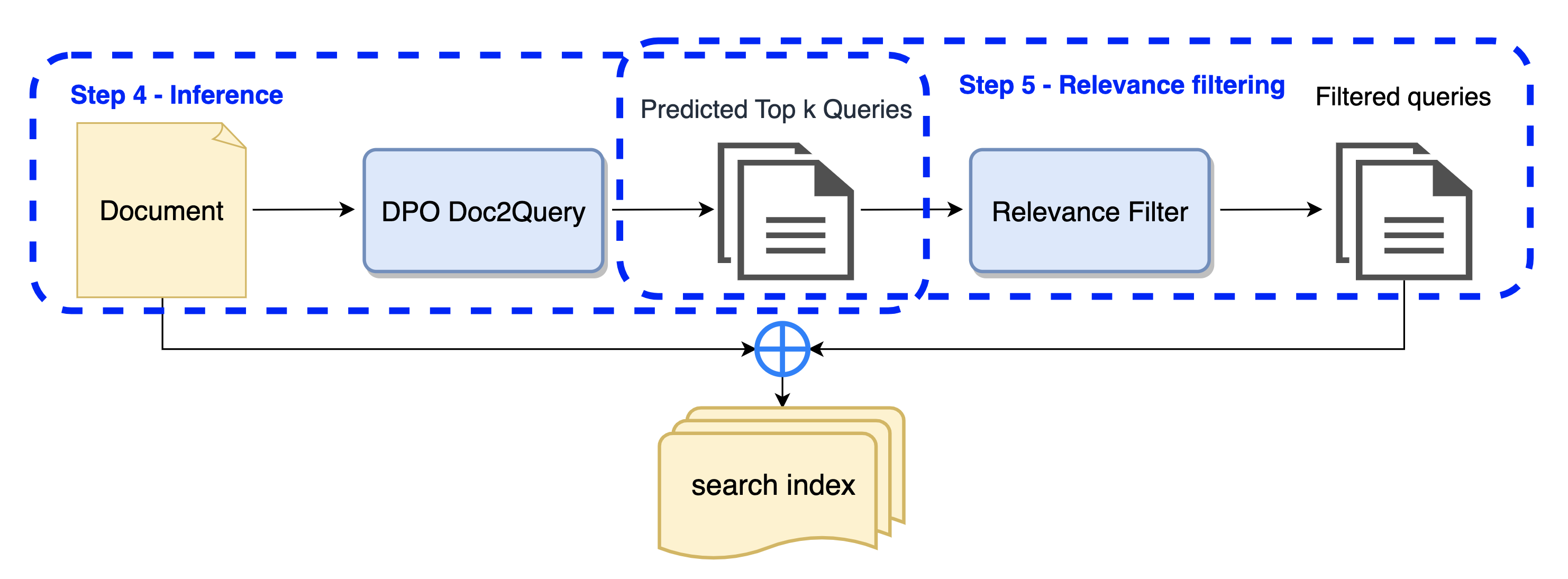}

    \caption{Implementation flow}

    \label{figure:dpo_inference}

\end{figure}

The vocabulary mismatch problem is a critical challenge for e-commmerce product search. This occurs when customers' search queries fail to align with the language used in product catalog descriptions. Bridging this gap is essential for effective product discovery and a positive online shopping journey. For example, a search for "smart watch for kid" may yield irrelevant results, as popular smart watches are usually recognized by their brands (e.g. apple, gabb, garmin) instead of explicitly mentioning "smart" in their product description. Due to the lack of match on "smart", popular brands are often overlooked. To bridge this linguistic gap, the approaches have been explored can be categorized into query expansion and document expansion. 

Query expansion methods focus on enriching customer search queries with relevant terms from the product catalog language, including supplementing the original query with synonyms, related terms, or category-specific keywords~\cite{rocchio71relevance, miller1995wordnet, lv2009comparative, liu2022query, wang2023query2doc}. In contrast, document expansion seeks to enrich product information with additional keywords drawn from customer language, with a broader, more customer-centric view of each product~\cite{nogueira2019document, nogueira2019doc2query, formal2021splade, gospodinov2023doc2query, li2024doc2token}.

Doc2Query~\cite{nogueira2019doc2query} is a popular document expansion technique which trains a sequence-to-sequence model, typically an encoder-decoder architecture with attention mechanisms, to generate potential queries relevant to the document. These generated queries are then appended to the document before indexing. While Doc2Query has proven remarkable effectiveness in improving lexical retrieval, it suffers from several drawbacks. A notable limitation is that it can generate repeated terms that are already present in the document \cite{li2024doc2token}. Efforts to encourage more diverse predictions by altering training data or loss function risk generating hallucinations or spurious content. For example, a Doc2Query model might generate queries containing unrelated tokens like "poster," "dvd," and "switch game" for a product titled "shadows of the damned | xbox 360."

Producing high-quality, novel, and relevant tokens remains challenging for Doc2Query models. To combat this, Doc2Query$--$~\cite{gospodinov2023doc2query} proposed a two-stage approach: first generating queries and then filtering them using a relevance model. Although this method has been shown to be effective and efficient online, it introduces the offline overhead of a generate-then-filter process. Critically, this approach doesn't solve the underlying issue: if the initial Doc2Query model produces low-quality results, the subsequent relevance filtering will yield only a limited set of relevant terms, ultimately reducing its impact.  A recent work \cite{mansour2024revisiting} explored replacing the "generate-then-filter" approach with a single generation phase using reinforcement learning. However, this method failed to produce diversified results and ultimately hurt retrieval performance.

This paper presents \textbf{Q}uery \textbf{G}eneration with \textbf{D}irect \textbf{P}reference \textbf{O}ptimization (QGDPO), a novel and effective approach for enhancing Doc2Query model performance by generating high-quality predictions. Drawing inspiration from Direct Preference Optimization (DPO) \cite{rafailov2024direct} in large language model (LLM) training, our method involves first training a sequence-to-sequence model to predict queries for a given document. Subsequently, we construct a relevance preference dataset by using a relevance model~\cite{reddy2022shopping,lin2024enhancing,vo2024knowledge,mehrdad2024large,shang2025knowledge,yang2026scaling}
as a proxy to distinguish winning and losing predictions. The triplets of product and associated winning and losing queries then serve as training data within a DPO framework, and a low learning rate is used to learn relevance preferences. Similar to \cite{gospodinov2023doc2query}, we apply a relevance filter to the QGDPO predictions before indexing. We deployed the model to production on Walmart.com, and it led to significant end-to-end improvements in relevance and engagement. We also explore LLM-based Doc2Query options such as Mistral 7B \cite{jiang2023mistral} to move beyond the popular T5 \cite{raffel2020exploring} architecture.

Our contributions are summarized as follows:

\begin{itemize}

    \item We introduce a novel QGDPO for document expansion enhancing the prediction quality for e-commerce search.

    \item Our proposed method (the combination of QGDPO and relevance filtering) has been successfully deployed on Walmart.com with significant metric lifts.

    \item We explore LLM-based models such as Mistral for Doc2Query on e-commerce data.

\end{itemize}

\section{Methodology}
\label{sec:methods}

As shown in Figure~\ref{figure:dpo_train} and Figure~\ref{figure:dpo_inference}, our proposed method builds upon the existing Doc2Query approach by incorporating two key enhancements: a DPO training phase after supervised fine-tuning (SFT), and a relevance filtering step after inference. Both DPO training and relevance filtering are guided by a relevance model that estimates how relevant a generated query is to the input product. This model is a multi-class classification model fine-tuned on human judgment data from Walmart, as described in \cite{lin2024enhancing}. It classifies query-product pairs into three categories: exact, substitute, and irrelevant, similar to Amazon ESCI definitions \cite{reddy2022shopping}. 

Our proposed approach consists of the following steps:

\textbf{Step 1 - SFT training}: For each product $D$, we fine-tune a base sequence-to-sequence model $f$ to generate a set of $n$ queries, $\mathcal{Q} = \{Q_{1}, \dots, Q_{n}\}$, i.e., $f: D \rightarrow \mathcal{Q}$. 

\textbf{Step 2 - Relevance preference}: 
We apply the fine-tuned model $f$ to all products in our training data.  We then use the relevance model to score each generated query $Q$ with respect to its input product $D$. This scoring function, $r$, outputs a vector of probabilities, i.e., $r: (D, Q) \rightarrow (r^E, r^S, r^I),$ the elements of which represent probabilities that query $Q$ for product $D$ belongs to the exact, substitute, and irrelevant classes, respectively. For each product $D$ in the training data, we construct DPO training triplets by sampling a winning query $Q_w$ from the set of queries labeled exact and a losing query $Q_l$ from the set of queries labeled irrelevant. Formally, the resulting DPO training data consists of triplets in the form $(D, Q_w, Q_l)$, where $r^E > \max(r^S, r^I)$ for $Q_w$ and $r^I > \max(r^E, r^S)$ for $Q_l$. 

\textbf{Step 3 - Preference fine-tuning}: A new model, denoted by $f^*$, is fine-tuned via DPO on the triplet data by minimizing the following loss function \cite{rafailov2024direct}:
\begin{align}
\label{loss_eq}
    \mathcal{L}_{DPO} = - \mathbb{E}_{(d, q_w, q_l)} \left[ \log \sigma \left( \beta \log \frac{f^*(q_w|d)}{f(q_w|d)}  -  \beta \log \frac{f^*(q_l|d)}{f(q_l|d)} \right) \right], 
\end{align} 
where $(d, q_w, q_l) $ are the triplets from $(D, Q_w, Q_l)$, $\sigma(z)$ is the sigmoid function, and $\beta$ is a hyperparameter that controls the strength of preference over the base SFT model $f$.

\textbf{Step 4 - Inference}: We apply the DPO fine-tuned model, $f^*$, to all products in our catalog to generate a new set of queries. 

\textbf{Step 5 - Relevance filtering}: To remove any noisy predictions that may still be present, we use the relevance model as a filter, retaining only the queries with an exact class probability exceeding a threshold. Eventually, the filtered queries are appended to the product and indexed for retrieval.

\begin{table*}[t]
\small
	\centering
	\vspace{-3mm}
	\begin{tabular}{c|c|cc|c}
	\hline
         \multirow{2}{*}{Model} & \multirow{2}{*}{Training description} & \multicolumn{2}{c|}{Lift on relevance metrics} &  \multirow{2}{*}{\# of novel tokens} \\
        \cline{3-4}
	   & & \% of exact & \% of irrelevant   &   \\
	\hline
        \hline
        \makecell{Doc2Query, baseline} & T5-base SFT & 0 & 0 & 6.0 \\
        \hline
        \makecell{QGDPO} & T5-base SFT + DPO & +8.07\% & -49.87\% & 5.0 \\        
        \hline
	\makecell{QGDPO$^2$} & T5-base SFT + DPO + DPO & +7.78\% & -50.40\% & 5.2 \\
	\hline 
        \makecell{Doc2Query$--$} & T5-base SFT & +1.49\% & -15.15\% & 5.8 \\
        \hline
        \makecell{QGDPO w/ relevance filtering, ours} & T5-base SFT + DPO & \textbf{+9.63\%} & \textbf{-64.48\%} & 4.9\\
        \hline 
	\end{tabular}
        \caption{Ablation studies for the top-5 query predictions on a small evaluation set of 1,000 prodcuts. Doc2Query$--$ in this table is the baseline Doc2Query with our relevance filter designed to fairly compare against our final model.} \label{tbl.dpo}
\vspace{-3mm}
\end{table*}
\hide{
\begin{table*}[t]
\small
	\centering
	\vspace{-3mm}
	\begin{tabular}{c|c|cc|cc}
	\hline
        \multirow{2}{*}{} & \multirow{2}{*}{Model description} & \multicolumn{2}{c|}{Small eval set} &  \multicolumn{2}{c}{Large eval set} \\
        \cline{3-6}
	   & & \% of exact & \% of irrelevant   & \% of exact & \% of irrelevant \\
	\hline
        \makecell{Doc2Query, baseline} & T5-base SFT & 0 & 0 & 0 & 0\\
        \hline
        \makecell{Doc2Query$--$} & T5-base SFT & +1.49\% & -15.15\% & - & - \\
        \hline
        \makecell{DPO Doc2Query} & T5-base SFT + DPO & +8.07\% & -49.87\% & +13.96\% &  -57.68\% \\
        \hline
	\makecell{DPO$^2$ Doc2Query } & T5-base SFT + DPO + DPO & +7.78\% & -50.40\% & +15.56\% & -66.29\%\\
	\hline 
        \makecell{DPO Doc2Query$--$, ours} & T5-base SFT + DPO & \textbf{+9.63\%} & \textbf{-64.48\%} & - & -\\
        \hline 
	\end{tabular}
        \caption{Ablation studies: Relevance metrics comparison for DPO models and baselines on the top-5 query predictions.} \label{tbl.dpo}
\vspace{-3mm}
\end{table*}
}

\begin{table}[t]
\small
	\centering
	\vspace{-3mm}
	\begin{tabular}{c|cc}
	\hline
	  Filtering Criterion & Precision & \# of novel tokens   \\
	\hline
        \hline
        %70\% percentile & 0.979 & 0.357 & 0.05 \\
        %\hline
        Recall=90\% & 0.887  & 4.20 \\
        \hline
        Recall=95\% & 0.866  & 4.90 \\
        \hline
        Recall=99\% & 0.838  & 5.71 \\
        \hline
        Recall=100\% & 0.824  & 6.00 \\
	\hline
	\end{tabular}
        \caption{Comparison of filtering strategies on our large evaluation set. The 95\% recall threshold provides a strong balance between precision and the number of novel tokens generated.} \label{tbl.threshold}
\vspace{-3mm}
\end{table}

\begin{table}[t]
\small
	\centering
	\vspace{-3mm}
	\begin{tabular}{c|cc}
	\hline
	  $\beta$ for DPO training & \% of exact & \% of irrelevant   \\
	\hline
        \hline
        \makecell{Doc2Query, baseline} & 0 & 0 \\
        \hline
        0.1 & \textbf{+3.26\%} & \textbf{-19.94\%} \\
        \hline
        0.5 & +3.23\% & -19.84\% \\
        \hline
        0.8 & +1.95\% & -12.02\%\\
        
	\hline
	\end{tabular}
        \caption{Relevance improvements over the Doc2Query baseline on large evaluation set by DPO training for 2 epochs with corresponding $\beta$s.} \label{tbl.beta}
\vspace{-3mm}
\end{table}

\begin{table}[t]
\small
	\centering
	\vspace{-3mm}
	\begin{tabular}{c|cc}
	\hline    Generation strategy  & \% of exact & \% of irrelevant \\
        \hline
        \makecell{Beam search\\
        (beams size=10, \\no repeat ngram size=2)}   & \textbf{92.43\%} & \textbf{2.89\%} \\
        \hline
        \makecell{Top-k sampling \\
        (k=10)}  & 87.73\% & 4.80\% \\
        \hline
	  \makecell{Top-k and top-p sampling \\ (k=10, p=0.95) } & 89.15\% & 4.20\%\\
	\hline
	\end{tabular}
        \caption{Relevance metrics for the top-5 predicted queries, generated by beam search, top-k sampling, and combined top-k/top-p sampling, evaluated on a large dataset.} \label{tbl.generation}
\vspace{-3mm}
\end{table}

\begin{table*}[ht]
\small
    \centering
    \begin{tabular}{p{1.9in}|p{4.8in}}
    \hline
        Doc2Query: & "small animal waterer", "\red{pig} waterer", "pigeon water dispenser", "\red{dog} waterer for small \red{dogs}", "pigeon water" \\ \hline
        QGDPO: & "pet auto feeder waterer", "auto pet feeder waterer", "automatic \red{dog} feeder waterer", "automatic water dispenser for pet", "auto waterer"  \\ \hline
        Queries dropped from Relevance Filtering: & "automatic \red{dog} feeder waterer" \\ \hline
    \end{tabular}
    \caption{Product title: 2pcs pet waterer automatic feeder waterer simple useful durable drinking kettle for pigeon parrot (small)}\label{tbl.example}
\end{table*}

\hide{
\begin{table*}[ht]
\small
    \centering
    \begin{tabular}{p{1in}|p{4.4in} | p{1in}}
    \hline
       Model & Prediction & Novel tokens
        \\ \hline
        Production Doc2Query: & "flower pot hanger", "black flower pot hanger", "black flower pot \red{holder}", "metal flower pot hook", "flower pot \red{holder}"  & metal, hanger, \red{holder} \\ \hline
        DPO Doc2query: & "hook for hanging plant", "plant hook", "hook to hang plant", "plant hook for hanging", "chain hook for hanging plant" & hanging, plant, hang \\   \hline
        DPO Doc2query w/ relevance filter: & "hook for hanging plant", "plant hook", "hook to hang plant", "plant hook for hanging" & hanging, plant, hang \\ \hline
    \end{tabular}
    \caption{Product title: black flower pot s-hook, 4 count, peerless chain company }\label{tbl.example}
\end{table*}
}

\section{Offline experiments}
\subsection{Datasets}
\label{subsec:data}
Similar to \cite{li2024doc2token}, we constructed our product-query dataset by sampling user engagement data from Walmart.com spanning two years. We then applied a series of preprocessing steps: relevance filtering, full match filtering, and price token filtering. In the relevance filtering step, we used the relevance model describe in Section \ref{sec:methods} to retain only the product-query pairs classified as exact-matches. Full match filtering removed the pairs where all query tokens were already present within the product information; this helps avoid generating redundant tokens. The price token filtering eliminated the price-related and promotional phrases (e.g., "below \$10", "on sale") from the training data to avoid predicting tokens that are offer-related and unrelated to the product itself. After those steps, we obtained a dataset of 64 million product-query pairs associated with 16 million products. We divided the dataset into train, validation, and test sets using an 8:1:1 ratio based on product to ensure no product overlap between the sets.

For offline evaluation, we used two datasets: a small set of 1000 randomly sampled products and a larger set of 1.6 million products from the test set. For each product, we generated the top 5 query predictions using beam search. These predictions were then assessed by human annotators on the three-class scale (exact, substituent, and irrelevant)  for the small set and by a relevance model for the large set. We report the percentage of exact-match predictions and irrelevant predictions. The relevance results are reported in terms of relative improvement compared to a baseline model. We primarily used the small evaluation set to compare different approaches, while the large set was used for hyperparameter tuning. Beyond relevance quality, we also track the number of novel tokens generated (i.e., tokens not present in the product title) to estimate the impact on retrieval.

\subsection{Experiment Setup}
% parameters 
For SFT, we fine-tuned the public T5-base model using eight Nvidia A-100 GPUs. We used a learning rate of 1e-4, a batch size of 64, a maximum input sequence length of 256, and maximum output sequence length of 32 for 5 epochs. For DPO training, we further fine-tuned the SFT model with a learning rate of 1e-6 on 10 million DPO triplets for 10 epochs. During model inference, we used a top-5 beam search strategy with a beam size of 10. The input to the model is a text of product information consisting of the product title and some attributes, such as brand, color, etc. 

\subsection{Offline results}
\label{subsec:results}

%For the experiments, we consider the production Doc2Query model used for document enrichment at Walmart.com as our baseline, which is based on a fine-tuned T5-small model. 

\subsubsection{Effectiveness of DPO training}
We measured the effect of DPO training using a small evaluation set of 1,000 products, comparing three models as detailed in Table \ref{tbl.dpo}. These included a baseline Doc2Query model, our QGDPO model (fine-tuned with one round of DPO), and QGDPO$^2$ (fine-tuned with two DPO rounds). We found that incorporating DPO training to explicitly learn relevance preferences led to significant improvements in relevance. We saw an 8.07\% increase in the prediction of relevant queries and a reduction of nearly 50\% in hallucinated or irrelevant query predictions. Despite these gains with a single DPO round, applying a second round (repeating Steps 2 and 3 from Section \ref{sec:methods} - the preference data creation and DPO fine-tuning steps) resulted in overfitting that prevented any further relevance improvement.

\subsubsection{Effectiveness of relevance filtering}
The relevance filtering cutoff is determined by the small evaluation set. We set the threshold to the level that ensures a 95\% recall rate for exact match products (details in Section \ref{subsec:discussion}). % TODO cite \cite{rossi2024relevance}
As shown in Table \ref{tbl.dpo}, this filtering step helped move the relevance metrics further. 
With the filtering on baseline Doc2Query model, we achieved a 1.49\% increase in exact-match predictions and a substantial 15.15\% reduction in undesirable predictions. Applying the same filtering technique to the QGDPO model produced similar gains, which indicated that the relevance improvement from QGDPO and relevance filtering are independent. Consequently, the combination of QGDPO and relevance filtering, achieved the best metrics on relevance. 

\subsection{Discussions}
\label{subsec:discussion}
\subsubsection{Impacts on novel tokens}
% shall we discuss using full index numbers and report in online analysis?
Given that the model is utilized for generating related terms for document for downstream lexical retrieval, there is a concern that improved relevance could lead to a reduction in the number of terms generated, and thus limiting its downstream impact. To empirically assess the effects on retrieval, we measured the number of novel tokens produced, as shown in Table \ref{tbl.dpo}. 
Our findings demonstrate that while DPO training and relevance filtering led to higher quality predictions, the number of novel tokens remained relatively stable, indicating that prediction diversity has not been compromised.

\subsubsection{Optimizing the relevance filtering threshold}
\begin{figure}[t]
    \centering
    \includegraphics[width=0.45\textwidth,height=0.3\textheight]{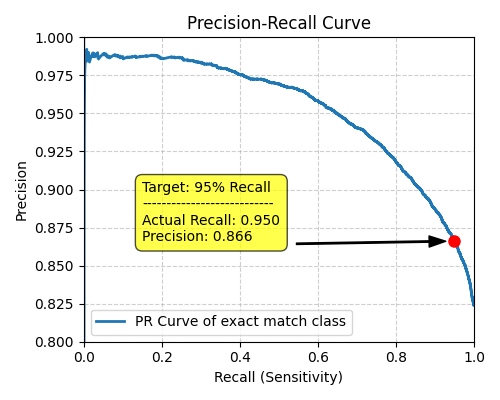}
    \caption{Precision recall curve of exact match class for relevance filter.}
    \label{figure:pr_curve}
\end{figure}
A critical step in our methodology is filtering generated queries to remove noise and retain only those with high relevance. To determine the optimal filtering threshold, we evaluated several strategies on our small evaluation set. Specifically, we apply a cutoff to the predicted probability of the exact-match class from our three-class relevance model. In Table \ref{tbl.threshold}, we measure the impact of different cutoffs on precision and the number of novel tokens generated per product. %First, we tested the 70th percentile cutoff used in prior work on passage ranking \cite{gospodinov2023doc2query}. This approach proved too stringent in our e-commerce domain, retaining only 35.7\% of relevant queries (recall=0.357, Figure \ref{figure:pr_curve}) with an average of just 0.05 novel tokens per product. This suggests that methods tuned for long-form passages do not directly transfer to the distinct characteristics of shorter e-commerce queries. 
We analyzed thresholds anchored to specific recall targets, visualizing the trade-off space in Figure \ref{figure:pr_curve}. While a 90\% recall threshold improved token generation, it was still overly restrictive. Conversely, a 99\% recall threshold was too loose, admitting a higher proportion of noisy queries and thus lowering precision (0.838). Based on this analysis, we adopted the threshold that achieves 95\% recall. This value represents an optimal compromise, ensuring the vast majority of relevant queries are included while maintaining high precision and effectively filtering out low-quality candidates that would otherwise degrade final retrieval performance.

\subsubsection{The choice of $\beta$}
To determine the optimal hyperparameter $\beta$ in Equation \ref{loss_eq}, we conducted DPO training for two epochs using three different $\beta$ values within the recommended range. We then evaluated the generated queries using our relevance model on the large evaluation set. The results, presented in Table \ref{tbl.beta}, showed that DPO training indeed improved prediction quality over the SFT Doc2Query model. However, we also observed that higher $\beta$ values led to increased hallucination. The optimal value for $\beta$ was determined to be 0.1.

\subsubsection{The choice of query generation strategies}
Previous research has provided insights into different decoding strategies for query generation. Nogueira et al. \cite{nogueira2019doc2query, nogueira2019document} observed that top-k sampling \cite{fan2018hierarchical} produces more effective queries than beam search. Conversely, Basnet et al. \cite{basnet2024deeperimpact} indicated that a combined top-k and top-p sampling approach surpasses pure top-k sampling in effectiveness.
We conducted experiments to evaluate these three decoding techniques (beam search, top-k sampling, and combined top-k/top-p sampling) on a large evaluation set. As detailed by the relevance metrics in Table~\ref{tbl.generation}, beam search notably outperformed the other methods, generating more exact-match queries and fewer irrelevant ones. While beam search is sometimes associated with output "degeneration" or lack of diversity \cite{holtzman2019curious}, its inherent stability is advantageous for our retrieval task. This consistency in query generation ensures more reliable and non-fluctuating retrieval results for a given input query.

\subsection{Examples}
Table~\ref{tbl.example} presents results for an example product. The production model Doc2Query generates hallucinated tokens, including those for unrelated animals (e.g., "pig", "dog"). The QGDPO model reduces some hallucinations by predicting "pet". Moreover, the relevance filter further removes irrelevant predictions (e.g. "dog") and guarantees that the predicted queries are relevant to the input product.

% Examples
%Table~\ref{tbl.example} showcases an example product along with its corresponding Doc2Query and Doc2Token predictions. The Doc2Query model produces the top 5 queries (consisting of 16 tokens in total), while the Doc2Token model generates the top 10 tokens. The Doc2Query model often produces queries containing tokens that are already present in the product information, while the Doc2Token model does not. %In contrast, all tokens produced by the Doc2Token model are relevant and absent from the product. Notably, the Doc2Token model is capable of predicting a Spanish word "salvavida" ("lifeguard" in English). This implies when a query "salvavida" comes, we can retrieve this product if we append Doc2Token predictions to the product for retrieval. %Queries in Spanish are commonly observed in e-commerce search.

\section{Online experiments}

% Implementation
% 1. Solr text matching 
% 2. results AB test  (online)
% 3. ndcg@10 for relevance 
We deployed our proposed approach in production through a pipeline outlined in Figure~\ref{figure:dpo_inference}. First, we performed model inference on all products in our catalog using cost-effective K80 GPUs. Then, a relevance model scored the predicted queries against their associated input items, and we filtered out queries with scores below a specified threshold. To improve efficiency, we only indexed predicted tokens not present in the product information. This offline inference process ran daily. For online usage, we created an additional text matching field based on QGDPO predictions in Solr~\cite{shahi2016apache}, an enterprise search platform built on Apache Lucene~\cite{lucene}, for enhancing text matching retrieval capabilities at Walmart.com.

For online testing, we compared the production Doc2Query model against our proposed QGDPO with relevance filtering. We evaluated the performance through a human relevance assessment on top-ranked products on the three-point scale (exact, substitute, irrelevant) and a two-week A/B test. We measured relevance using NDCG@5 and NDCG@10 and customer engagement through search session Add-to-Cart (ATC) rate and Gross Merchandise Value (GMV) in Table~\ref{tbl.online}. Relevance assessment showed statistically significant improvements in both NDCG@5 and NDCG@10. The A/B test revealed a statistically significant increase of 0.36\% in search session ATC rate, alongside a positive trend in GMV. These results demonstrate that our model enhances relevance and positively impacts user engagement.

\begin{table}
\small
\begin{center}
\begin{tabular}{c|c|c}
 \hline
 Metric & Lift & P-value\\ 
 \hline
  \hline
  NDCG@5 & \textbf{+0.93\%}  & 0.00\\ 
  \hline
  NDCG@10 & \textbf{+0.86\%} & 0.01\\ 
  \hline
  Search Session ATC Rate  & \textbf{+0.36\%}  & 0.01 \\
  \hline
%  Orders & +0.21\%   & 0.44\\
%  \hline
  GMV & +0.17\% & 0.36 \\
  \hline
\end{tabular}
\end{center}
\caption{Online experiment results. Statistically significant values are bold.}
\label{tbl.online}
\end{table}

%During our A/B testing of the Doc2Token feature, we noted a notable increase in revenue by +0.17\%. In terms of relevance, our results show a +0.49\% improvement in NDCG@10 (0.066) within the impacted traffic. Our analysis suggests that reintroducing relevant items into the search results enhances quality and aids customers in finding what they are searching for. 

\section{Mistral experiments}
\begin{table}[t]
\small
	\centering
	\vspace{-3mm}
	\begin{tabular}{c|cc}
	\hline
	 SFT models  & \% of exact & \% of irrelevant   \\
	\hline
        \hline
        T5-base & 0 & 0 \% \\
        \hline
        Mistral 7B & +2.00\% & -22.84\% \\ 
	\hline
	\end{tabular}
        \caption{Relevance improvements of Mistral Doc2Query model over T5-base model on large evaluation set} \label{tbl.mistral}
\vspace{-3mm}
\end{table}
Given the demonstrated potential of LLMs in natural language processing tasks, such as document expansion~\cite{basnet2024deeperimpact}, we investigated using the Mistral-7B model as a replacement for T5-base in Doc2Query tasks. We fine-tuned Mistral-7B-v0.1\cite{jiang2023mistral} using Low-Rank Adaption (LoRA) \cite{hu2021lora} with a rank of 256, alpha of 64 and a learning rate of 1e-4. Training was conducted over 3 epochs with a batch size of 4 on 8 Nvidia A100 GPUs. The Doc2Query T5-base model was used as a baseline for comparison.

The models were evaluated on the large evaluation set, as shown in Table \ref{tbl.mistral}. Mistral-7B achieved a significantly higher exact match rate of 2.00\% compared to T5-base model. It also significantly reduced irrelevant queries by 22.84\%. Mistral-7B's ability to leverage its broader world knowledge, likely derived from pre-training on internet-scale data, proved particularly beneficial for products related to media or toys. For example, given the product title "Transformers Revenge Of The Fallen Deluxe Class Wave 7 Arcee Action Figure", which is a character from Transformers film series, Mistral-7B generated queries like "transformers arcee toy", "deluxe class transformers figure", "autobot arcee action figure", "revenge of the fallen toy" and "transformers motorcycle figure". By inferring that Arcee is a motorcycle, the model introduced novel and relevant tokens such as "toy", "autobot", and "motorcycle". This enhanced product discoverability, enabling retrieval via searches like "motorcycle toys", which might be missed by T5 models with less world knowledge.

Despite its larger size of 7 billion parameters, Mistral-7B provided notable improvements over the T5-base model, which has 220 million parameters. This suggests that larger models can generate more accurate and relevant queries. However, their higher computational and memory demands should be considered for practical implementation.

\section{Related Work}

\textbf{Preference Optimization Methods.} To make language models generate better content, researchers use methods based on rewards of human preference. A common approach is Reinforcement Learning from Human Feedback (RLHF) \cite{christiano2017deep,ouyang2022training}, often using algorithms like Proximal Policy Optimization (PPO) \cite{schulman2017proximal}. PPO typically relies on a separate "reward model" trained to score outputs based on these preferences. More recently, Direct Preference Optimization (DPO) \cite{rafailov2024direct} offers a simpler way. DPO trains the main model directly with preference pairs (chosen vs. rejected) without needing a separate reward model, thereby simplifying the training pipeline and directly incorporating preference signals. This makes it more direct and often more stable. Distinct from these, Group Relative Policy Optimization (GRPO) \cite{ramesh2024group,shao2024deepseekmath} takes a different approach by formulating an objective function that minimizes the loss for the worst-case group. This design explicitly aims to ensure fairness by prioritizing disadvantaged groups during the optimization process. Our QGDPO uses DPO because it's straightforward and effectively learns from our "good query" vs. "bad query" examples, guiding the generator to produce better results. Our QGDPO method leverages DPO due to its directness and effectiveness in learning from the pairwise (document, winning query, losing query) preference triplets constructed via our relevance model, which aligns well with our goal of improving overall query relevance based on individual document-query pairs rather than specific group fairness objectives.

\textbf{Document expansion.} Document expansion enriches document representations to better match user queries. Doc2Query (\cite{nogueira2019document,nogueira2019doc2query}) is a prominent technique that uses sequence-to-sequence models to generate potential queries for a document. While effective, Doc2Query can produce irrelevant "hallucinations" or repetitive content. In contrast, Doc2Token \cite{li2024doc2token} proposed predicting missing relevant tokens for a document rather than entire queries, aiming to reduce redundancy and improve the novelty of added terms effectively. Efforts to improve generation quality include post-generation filtering, as seen in Doc2Query$--$ \cite{gospodinov2023doc2query}, which removes low-quality queries using a relevance model. While helpful, this doesn't fix the underlying generator.  More recently, preference alignment and reinforcement learning (RL) techniques, such as PPO, DPO, and GRPO, have been explored to refine query generation and document summarization for search relevance~\cite{mansour2024revisiting,yadav2025generating}. However, applying standard RL algorithms can severely penalize query diversity and hurt overall retrieval performance. This highlights the difficulty in balancing relevance and diversity with standard RL techniques for this task. Our approach, QGDPO, also uses a relevance filter, but critically, we first improve the generator itself. 

\textbf{Large Language Models for Expansion.}
The advancement of LLMs has led to their exploration in various information retrieval tasks. For document expansion, Basnet et al. \cite{basnet2024deeperimpact} explored using LLMs to generate expansions specifically for learned sparse retrieval mechanisms. Wang et al. \cite{wang2023query2doc} explored using LLMs to generate relevant queries given a document context, which, although framed as query expansion in their work, shares the core generative task with Doc2Query.Our work also investigates using more powerful LLMs like Mistral-7B \cite{jiang2023mistral} as a base for Doc2Query, comparing its performance against traditional T5-based architectures \cite{raffel2020exploring} in the product search domain to assess the benefits of increased model scale for generating relevant queries.

In essence, QGDPO advances upon Doc2Query by integrating DPO to directly enhance the quality of generated queries, complemented by a relevance filter. This distinguishes our method from prior work that either relied solely on post-hoc filtering or reinforcement learning techniques that showed limitations in this domain.
\section{Conclusions}
This work introduces a novel approach, QGDPO with a relevance filter, for generating high-quality queries for document expansion in product retrieval. Our experiments show that this method achieves superior relevance performance compared to the Doc2Query baseline. We also investigate the use of LLMs for document expansion tasks. The proposed feature has been successfully deployed and validated through online testing, which resulted in significant improvements in both relevance and user engagement.

\bibliographystyle{ACM-Reference-Format}
\bibliography{sigproc,references_trimmed}

\begin{screenonly}
\newpage
\appendix

\end{screenonly}

\end{document}